\documentclass[conference]{IEEEtran}

\usepackage[noadjust]{cite}
\usepackage{graphicx}
\usepackage{graphicx,color,psfrag,subeqnarray}
\usepackage{amsmath}
\usepackage{amssymb}
\usepackage{amsthm,amssymb}
\usepackage{color}
\usepackage{listings}
\usepackage{multirow}
\usepackage{bm,array}
\usepackage{listings}
\usepackage{caption}
\usepackage{subcaption}
\usepackage{stfloats}
\usepackage{float} 
\usepackage{subfloat}
\usepackage{bigstrut}

\newtheorem{theorem}{Theorem}[section]

\newtheorem{proposition}[theorem]{Proposition}

\begin{document}

\title{A Statistical Modelling and Analysis of Residential Electric Vehicles' Charging Demand in Smart Grids}

\author{\IEEEauthorblockN{Farshad Rassaei, Wee-Seng Soh and Kee-Chaing Chua \\}
\IEEEauthorblockA{Department of Electrical and Computer Engineering\\
National University of Singapore, Singapore\\
Email: \{f.rassaei, weeseng, eleckc\}@nus.edu.sg}
}
\maketitle

\begin{abstract}

Electric vehicles (EVs) add significant load on the power grid as they become widespread. The characteristics of this extra load follow the patterns of people's driving behaviours. In particular, random parameters such as arrival time and charging time of the vehicles determine their expected charging demand profile from the power grid. In this paper, we first present a model for uncoordinated charging power demand of EVs based on a stochastic process and accordingly we characterize an EV's expected daily power demand profile. Next, we illustrate it for different charging time distributions through simulations. This gives us useful insights into the long-term planning for upgrading power systems' infrastructure to accommodate EVs. Then, we incorporate departure time as another random variable into this modelling and introduce an autonomous demand response (DR) technique to manage the EVs' charging demand. Our results show that, it is possible to accommodate a large number of EVs and achieve the same peak-to-average ratio (PAR) in daily aggregated power consumption of the grid as when there is no EV in the system. This peak value can be decreased further significantly when we add vehicle-to-grid (V2G) capability in the system. 
\end{abstract}

\IEEEpeerreviewmaketitle

\section{Introduction}
Normally, the daily residential power demand profile has a significant peak-to-average ratio (PAR) that can potentially reduce the power grids' efficiency and incur exorbitant costs for developing the power grid's infrastructure, i.e., increasing the power generation, transmission, and distribution capacity of the grid. This extra capacity is just to serve the power demand of the users during transient peak-time periods. Hence, obviating this drawback has motivated intensive research on strategies that can utilize the existing grid more efficiently so that more consumers can be accommodated and served without developing new costly infrastructure. The main objective of these strategies is to make the demand responsive \cite{mohsenian-rad_autonomous_2010}.  

Demand response (DR) is predicted to become even more crucial as the use of new electricity-hungry appliances such as plug-in electric vehicles (PEVs) is becoming more widespread. Typically, on charging mode, they can double the average dwelling's energy consumption, with current electric vehicles (EVs) consuming 0.25-0.35 kWh of energy for one mile of driving \cite{van_roy_apartment_2014}. Hence, EVs' uncoordinated charging, i.e., the battery of the vehicle either starts charging as soon as plugged in or after a user-defined delay, can significantly exacerbate the already high PAR. 

Although it makes sense to envisage the number of electric cars increasing, it is hard to see that the electricity infrastructure capacity growing with the same rate concurrently. Thus, the ramification of introducing a large number of EVs into the grid has become an important avenue for research in recent years in the context of smart grid \cite{shao_challenges_2009}. First, we need to ask how uncoordinated charging can affect the existing power grid. Next, we need to ask, how we can satisfy this charging demand efficiently when we have information exchange capability and intelligence in a smart grid.   

There are several prior literature on modelling the impact of uncoordinated charging of EVs. However, most of them require much detailed information about passenger car travel behaviour, e.g., \cite{grahn_phev_2014} and \cite{lee_synthesis_2011}. Not only are the models mostly complicated and very test-oriented, but the sensitivity of the EVs' charging load to different parameters is not also clear. Moreover, most of previous models do not provide expected daily power demand resulted from EVs, particularly when EVs are charged in households rather than in charging stations. For instance, \cite{Spatial} provides a spatial and temporal model of electric vehicles charging demand for fast charging stations situated around highway exits based on known traffic data. In \cite{grahn_phev_2014}, a utilization model is proposed based on type-of-trip. The authors in \cite{li_modeling_2012} have used random simulation and statistical analysis to fit a distribution for the overall charging demand of EVs mainly for probabilistic power flow calculations. In \cite{alizadeh_scalable_2014}, the daily load profile is modelled by using queuing theory and the approach is suitable mainly for accurate short-time load forecasting.

Furthermore, since EVs are considered as the main component of the residential flexible electricity demand, numerous researches have been carried out for EVs' DR, e.g., \cite{clement_coordinated_2009} and \cite{Clement}. In \cite{kim_bidirectional_2013}, the authors have investigated the impact of vehicle-to-grid (V2G) energy delivery on the social welfare. Additionally, EVs' storage capacity can be used for improving the power grid's reliability, e.g., in terms of frequency control \cite{moghadam_randomized_2013}. But, the main drawback in most of these works is that they do not consider the inherent randomness of this demand in the first place. 

In this paper, we first present a model for uncoordinated charging power demand of a typical EV by formulating it as a stochastic process based on the arrival time and charging time of the vehicle. Here, the charging is taken place at users' homes and we treat EVs the same as other household electrical appliances. We then characterize an EV's expected daily power demand profile and illustrate it for different charging time distributions through simulations. The power demand profile of EVs gives us useful insights into the long-term planning for upgrading power systems' infrastructure to accommodate a large number of EVs. Next, we incorporate departure time as another random variable into this modelling and introduce an autonomous demand response (DR) technique to manage the EVs' charging demand mainly to flatten the daily aggregated power demand profile. Our results show that, it is possible to accommodate a large number of EVs and achieve the same peak-to-average ratio (PAR)  as when there is no EV in the system.


\section{System Model} \label{SM}

\begin{figure}
	\centering
	\includegraphics[width=0.8\columnwidth]{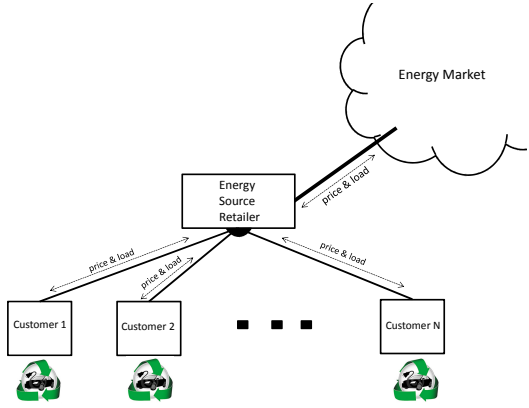} 
	\caption{Basic model of a smart energy system comprised of multiple load customers which share one energy source retailer or an aggregator.} \vspace{-1.5em}
	\label{f1}
\end{figure}

Fig. \ref{f1} represents a basic power system model where multiple energy customers share one energy source retailer or an aggregator. We assume that the consumers' total load consists of two different types of load; normal inflexible household load which needs \textit{on-demand} power supply, e.g. air conditioning, lighting, cooking and refrigerator, and EV as a flexible load. 
 
\begin{figure}
  	\centering
  	\includegraphics[width=0.8\columnwidth]{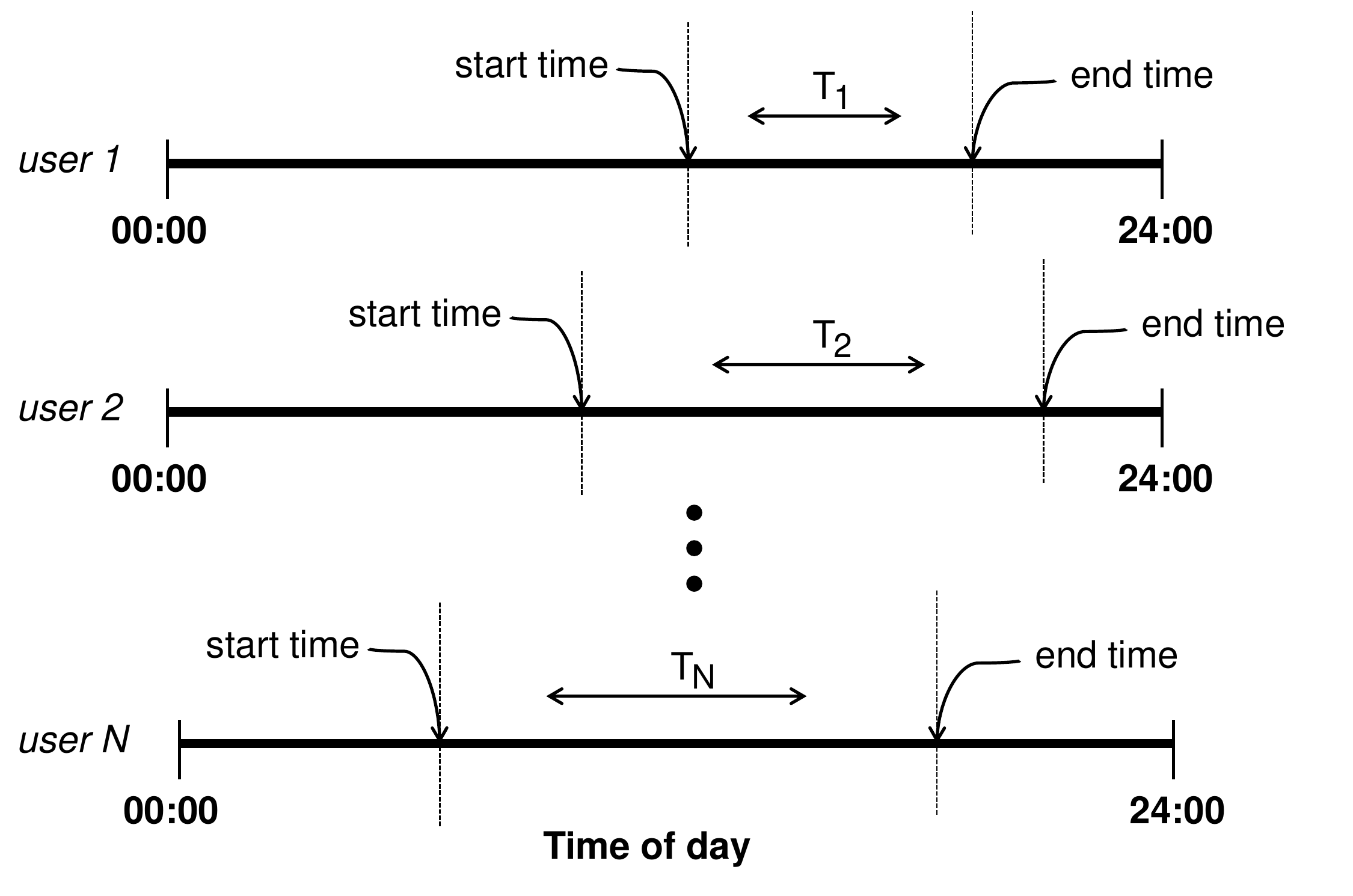}
  	\caption{Time setting for accomplishing a certain job on an appliance for different users during a day.}
  	\label{f4}
  \end{figure}

Fig. \ref{f4} shows the concept of power demand flexibility for an EV for different users. The charging process of user $n$, $n\in \{1,...N\}$, may take time $T_n$ to be completed. In addition to the start time, the users set the deadline by which this job should be accomplished. In this case, we may recognize the following three random variables for an EV's charging process:

\begin{itemize}

\item \textbf{Start Time} shows the time when the EV connects to the grid and delivering energy can potentially start.  

\item \textbf{Charging Time} which generally differs from one user to another according to the vehicle's driven distance.  

\item \textbf{End Time} represents the deadline specified by the user for accomplishing the charging process.   

\end{itemize}

Therefore, in general, we can formulate the \textit{uncoordinated} charging power demand for an EV as follows:  

	\begin{equation}
	x(t) \triangleq
	\begin{cases}
	a  &   t_0\leq t < t_0+T  \\
	0       & otherwise \\
	\end{cases}
	\label{xt}
	\end{equation}

\noindent{where we consider instantaneous power consumption as the constant $a$ and assume that power consumption in standby mode is negligible. Additionally, $T$ and $t_0$ denote the charging time and the start time, respectively. These parameters are random in general. Here, we assume $t_0$ and $T$ have independent PDFs that can be found from empirical data. For example, for the arrival time $t_0$ a Gaussian distribution is proposed in \cite{lee_stochastic_2012}. Here, we are mainly interested in knowing the daily power consumption profiles, i.e., the power consumption behaviour throughout a typical 24-hour day. Therefore, we calculate (\ref{xt}) in modulo 24-hours and then project the results onto a 24-hour day. In this case, some realizations of the stochastic process defined in (\ref{xt}) can be displayed as shown in Fig. \ref{f7}. This figure shows (\ref{xt}) for ten different users in a bar graph with one hour time granularity.

  \begin{figure}
  	\centering
  	\includegraphics[width=0.8\columnwidth]{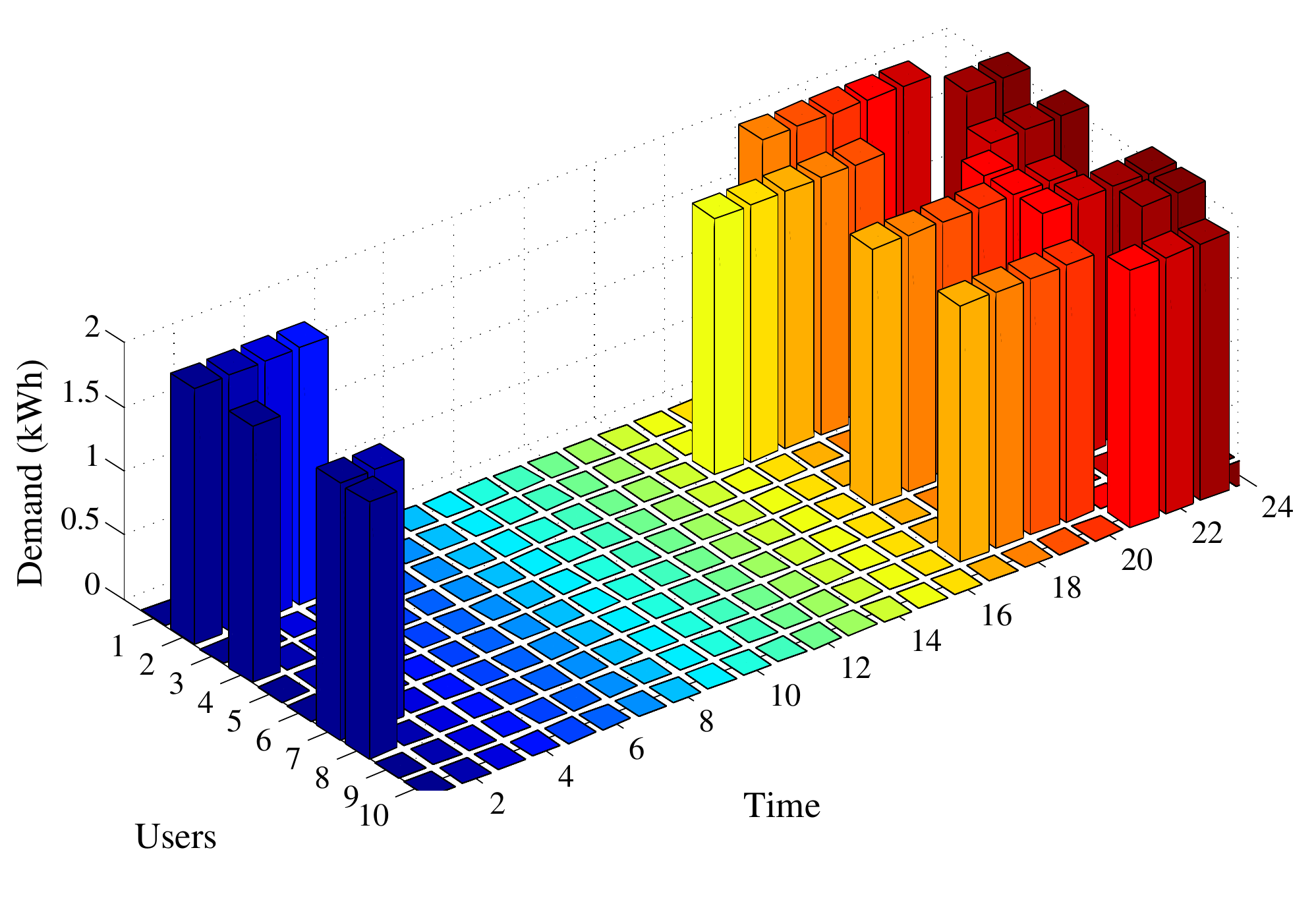}\vspace{-1em}
  	\caption{Some realizations of the stochastic process defined in (\ref{xt}) in modulo 24-hours.} \vspace{-1em}
  	\label{f7}
  \end{figure}

\section{Statistical Analysis} \label{Sta}

In this section, using the aforementioned definition of $x(t)$, we calculate $\mathbb{E}[x(t)]$ which represents the expected value of uncoordinated charging power consumption for an EV. This expectation can be expressed by the following proposition for EVs (refer to the appendix for the proof). 

\begin{proposition}
Given $f_{t_0}(\cdot)$ and $f_T(\cdot)$ as the PDFs of the independent random variables arrival time $t_0$ and charging time $T$ for an EV, the expected uncoordinated charging power demand can be expressed as:

\begin{gather}
 \mathbb{E}[x(t)]=a\times \big( F_{t_0}(t) \ast [\delta(t)-f_T(t)]  \big)
\label{EX}
\end{gather}
\noindent{in which, $\ast$ shows the convolution operation and $\delta(t)$ is the Dirac delta function. Also, $F(\cdot)$ represents the cumulative distribution function (CDF).} 

\label{pro}
\end{proposition}

In addition, from (\ref{EX}), the expected time of maximum power consumption can be found from the following equation:
\vspace{-0.4em}
\begin{gather}
f_{t_0}(t_{max})= f_{t_0}(t_{max}) \ast f_T(t_{max}).
\label{EXmax}
\end{gather}
We can calculate (\ref{EX}) for any given distribution analytically or numerically. In the sequel, we adopt different distributions for the EV's charging time $T$ following some available empirical research data in the literature, as shown in Fig. \ref{besuni}, to study the corresponding results of (\ref{EX}). We investigate four cases for the distribution of $T$, namely, the uniform, exponential, Gaussian with positive support, and Rician distributions. These distributions have different degrees of freedom (DoF) and all of them support $T$ over $[0,+\infty)$:
\begin{itemize}
\item{\textbf{T: Uniform} } In this case, we consider $T$ to have uniform distribution over the interval $[c,d)$. Then, $\mathbb{E}[x(t)]$ can be simply derived. 
\end{itemize}

Assuming $t_0$ has a normal distribution with mean $\mu$ and variance $\sigma^2$ and $T$ has a uniform distribution over the interval $[c,d)$, $0\leq c < d$, the expected uncoordinated charging power demand is given by:  
\begin{gather}
\nonumber \mathbb{E}[x(t)]=a\times 
\bigg[ 1-\mathbf{Q}(\frac{t-\mu}{\sigma})+\frac{\sigma}{d-c}
( c'\mathbf{Q}(c')\\
-d'\mathbf{Q}(d')+f(d')-f(c')+d'-c' ) \bigg]
\end{gather}
\label{prou}
where $c'=\frac{t-c-\mu}{\sigma}$, $d'=\frac{t-d-\mu}{\sigma}$, and 

\begin{equation}
\mathbf{Q}(x)=\frac{1}{\sqrt{2\pi}}\int\limits_x^\infty \exp (-\frac{u^2}{2}) du,
\end{equation} 
 \vspace{-0.5em}
\begin{equation}
f(x)=\frac{\exp(-\frac{x^2}{2})}{\sqrt{2\pi}}.
\end{equation}

\begin{figure}
      \centering
      \includegraphics[width=0.8\columnwidth]{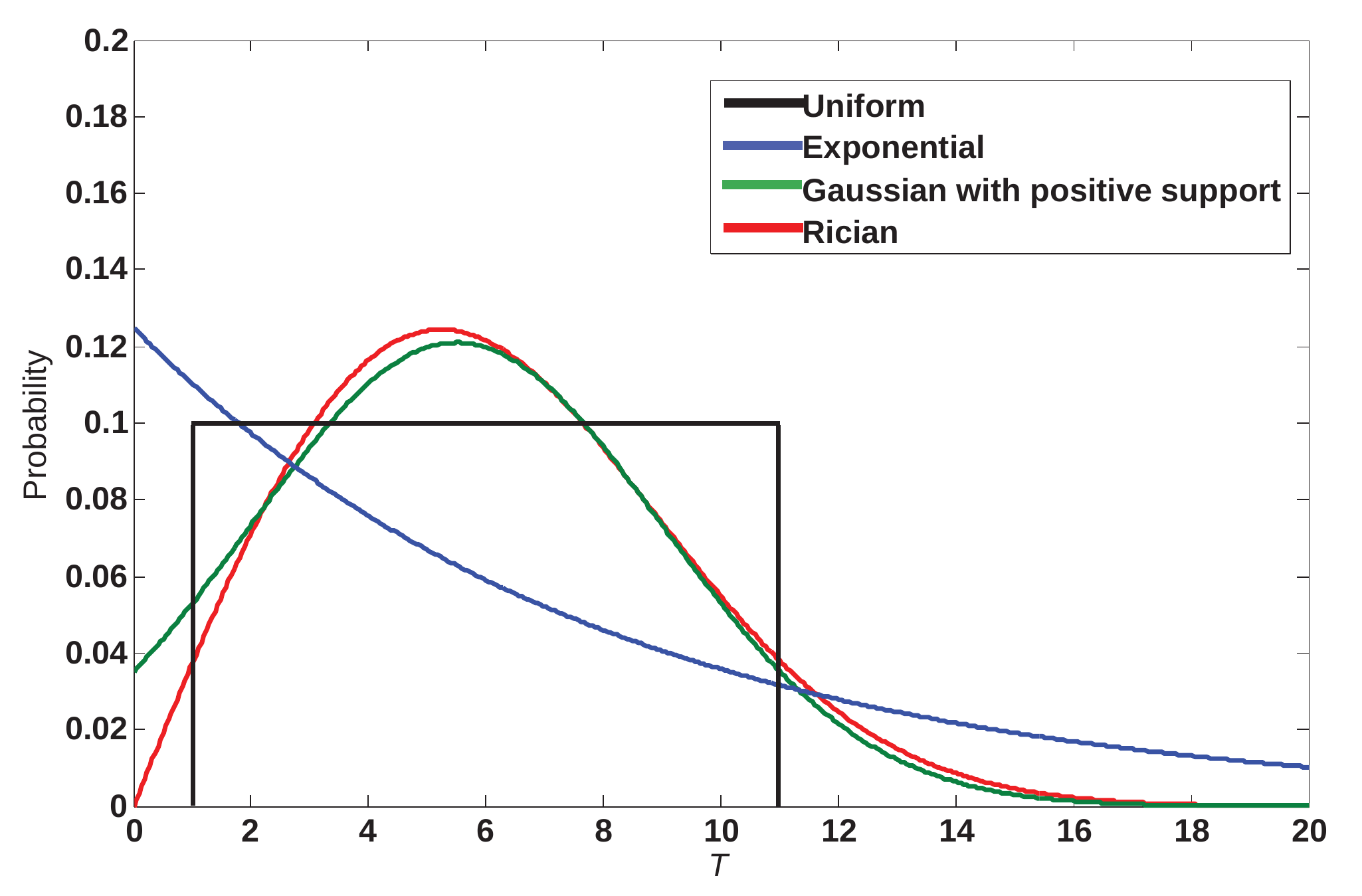}\vspace{-1em}
      \caption{Uniform, exponential, Gaussian with positive support and Rician distributions for $T$.} \vspace{-1.5em}
      \label{besuni}
\end{figure}

\begin{itemize}

\item{\textbf{T: Exponential}} The driven distance and hence the charging time of an EV can be modelled by an exponential distribution \cite{liang_stochastic_2014}. For an exponentially distributed $T$ with mean $\lambda^{-1}$, we have the following PDF: 

\begin{gather}
f_T(T)=\lambda\exp(-\lambda T).
\label{pexp}
\end{gather}

\item{\textbf{T: Gaussian}} When $T$ has a Gaussian PDF with positive support as shown in Fig. \ref{besuni}, $T$ has the following distribution function: 
\vspace{-0.5em}
\begin{gather}
 f_T(T)=N(T;\mu,\sigma^2|0\leq T < \infty),\\
=\frac{1}{\mathbf{Q}(\frac{-\mu}{\sigma})\sqrt{2\pi \sigma^2}}\exp (-\frac{(T-\mu)^2}{2\sigma^2}),\, \,\,\, 0\leq T < \infty. 
\label{pgauss}
\end{gather}

\item{\textbf{T: Rician}} Finally, we consider a Rician PDF for $T$ having the following form:

\begin{equation}
f(T|\nu , \sigma)=\frac{T}{\sigma^2}\exp(-\frac{(T^2+\sigma^2)}{2\sigma^2})I_0(\frac{T\nu}{\sigma^2}),
\label{rice}
\end{equation}
where $\nu \geq 0$ and ${\sigma} \geq 0$  present the noncentrality parameter and scale parameter, respectively. $I_0(\cdot)$ is the modified Bessel function of the first kind with order zero. 
\end{itemize}

\section{Demand Response} \label{DR}
In this section, we apply an autonomous DR approach to manage the EVs' stochastic charging demands mainly to flatten the daily aggregated power demand profile. Let $N$ denote the number of users that share an energy retailer or an aggregator according to Fig. \ref{f1}. Each customer $n'$s load at time slot $t$ can be denoted by:

\begin{align}
l^{t}_{n}=l^{t}_{EV,n}+l^{t}_{A,n}, \quad t\in\mathbb{T}\triangleq\{1,\dotsc,H\} 
\end{align}
where $H$ is the scheduling horizon and, $l^{t}_{EV,n}$ and $l^{t}_{A,n}$ represent the EV's load and the overall load from the household appliances at time slot $t$, respectively. Hence, the time-varying load profile for user $n$ over the scheduling horizon is denoted by the following vector: 

\begin{equation}
\textbf{l}_{n}\triangleq [l^{1}_{n},\dotsc,l^{H}_{n}]^T=\textbf{l}_{EV,n}^T+\textbf{l}_{A,n}^T.
\end{equation}

Here, without loss of generality, we assume a daily scheduling horizon. Each user tries to minimize the correlation between its EV charging demand profile and the aggregated demand profile from the other users in the system and its own inflexible power demand as expressed in the following problem:  

\begin{equation}
\begin{aligned}
& \underset{\textbf{l}_{EV,n} \in \textbf{l}^{P}_{EV,n} }{\text{minimize}}
& & { < \textbf{l}_{EV,n},\textbf{l}_{A,n} + \sum_{\substack{i \in {N} \\
i \neq n}}  (\textbf{l}_{EV,i}+\textbf{l}_{A,i}) > }  \\
\end{aligned}
\label{cor}
\end{equation}

\noindent where $<\textbf{x},\textbf{y}>$ shows the inner product between vectors $\textbf{x}$ and $\textbf{y}$ representing their correlation and $\textbf{l}^{P}_{EV,n}$ is:
 
\begin{align}
\textbf{l}^{P}_{EV,n} =\bigg\{ \textbf{l}_{EV,n}\mid \sum_{t=\alpha_{n}}^{\beta_{n}} l^{t}_{EV,n}=E_{EV,n}; \nonumber \\ |l^{t}_{EV,n}| \leq p_{max}; \quad l^{t}_{EV,n}=0, \quad \forall t\setminus{T}^P_{EV,n} \bigg\} .
\end{align}

\noindent where, $E_{EV,n}$ is the $n^{th}$ user's required energy to charge its EV which determines the charging time while $\alpha_{n}$ and $\beta_{n}$ present the arrival time and departure time of the EV which are random as discussed in the previous section. Furthermore, ${T}^P_{EV,n}$ represents the permissible charging time set by the user and $|l^{t}_{EV,n}| \leq p_{max}$ limits the maximum power that can be delivered to/from the EV. The term $\sum_{\substack{i \in {N} \\
i \neq n}}  (\textbf{l}_{EV,i}+\textbf{l}_{A,i})$ is the state of the aggregated load profile from the other $N-1$ users in the system. This state information can be provided from the aggregator to each user. Then, this problem is solved iteratively by all the $N$ users in the system and the aggregator updates the state information after each iteration.

\section{Simulation Results} \label{SR}

In this section, we consider a Gaussian distribution for the random variable $t_0$ as the arrival time with $\mu=19$ and $\sigma^2 = 10$ inspired from \cite{lee_stochastic_2012}. Furthermore, we consider four cases for the distribution of the random variable $T$ as described in section \ref{Sta}. First, we consider $T$ to have a uniform distribution over the interval $[1,11]$. Second, we assume $T$ to be exponentially distributed with mean $\mu=6$. Third, we assume $T$ to be Gaussian distributed with positive support as presented in (\ref{pgauss}). In this case, we use the well-known \textit{accept-reject} approach to generate the random values. 
Finally, we consider a Rician distribution for $T$. In all cases (except for the exponential distribution), we set the parameters of the distributions such that they all have the same mean and variance. However, for the exponential distribution case, we can only set either its mean or variance to be the same as that of the others since this distribution has just one DoF. Based on an average 0.25 kWh energy consumption for each mile of driving, we set all the parameters in (\ref{xt}).

\begin{figure} 
	\centering
		\includegraphics[width=0.8\columnwidth]{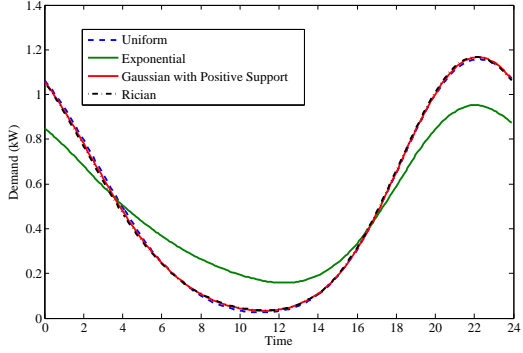}\vspace{-0.5 em}
	\caption[]{An EV's expected daily power demand profile for different distributions of charging time $T$.} \vspace{-1em} 
	\label{com}
\end{figure}

\begin{figure}[t!]  \centering
 \begin{subfigure}[b]{\columnwidth}\centering
 \includegraphics[height=2.018in]{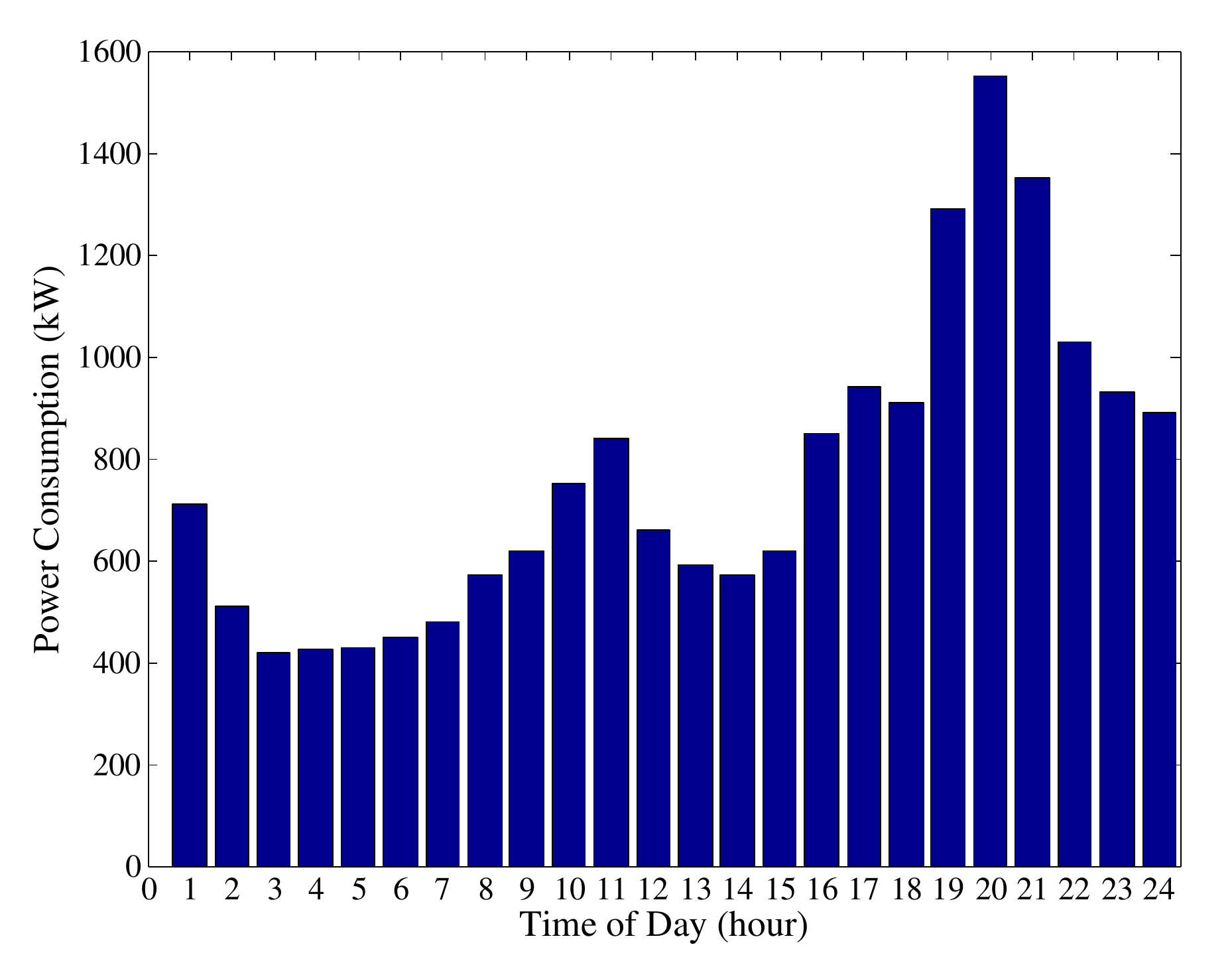}
 \vspace{-0.4 em} 
 \caption{Aggregated inflexible daily power demand profile}
 \label{inflex}
 \end{subfigure} 
  \begin{subfigure}[b]{\columnwidth}\centering
  \includegraphics[height=2.018in]{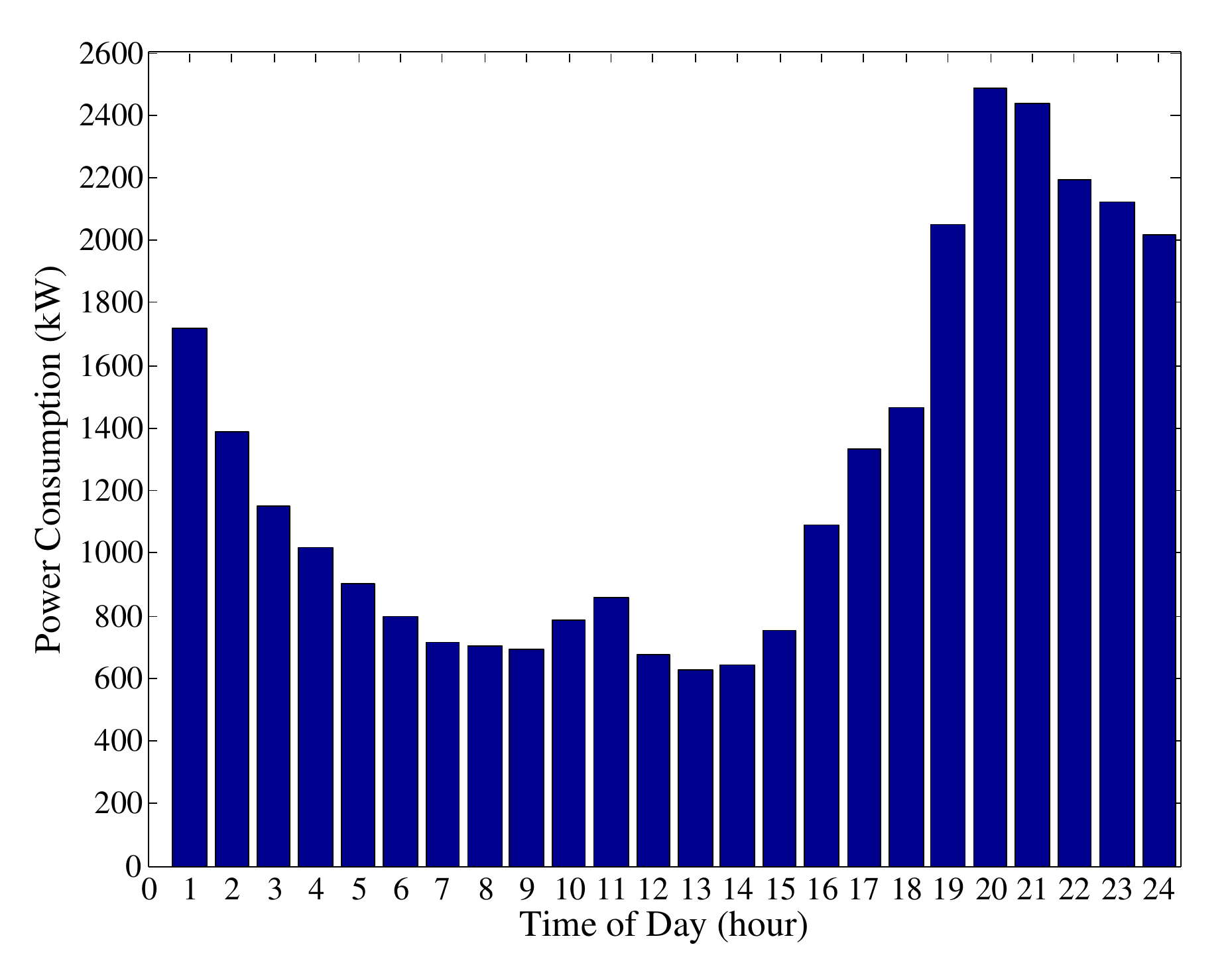}
  \vspace{-0.4 em} 
  \caption{Aggregated total daily power demand profile}
  \label{inflex&EV}
  \end{subfigure}
     \begin{subfigure}[b]{\columnwidth}\centering
    \includegraphics[height=2.018in]{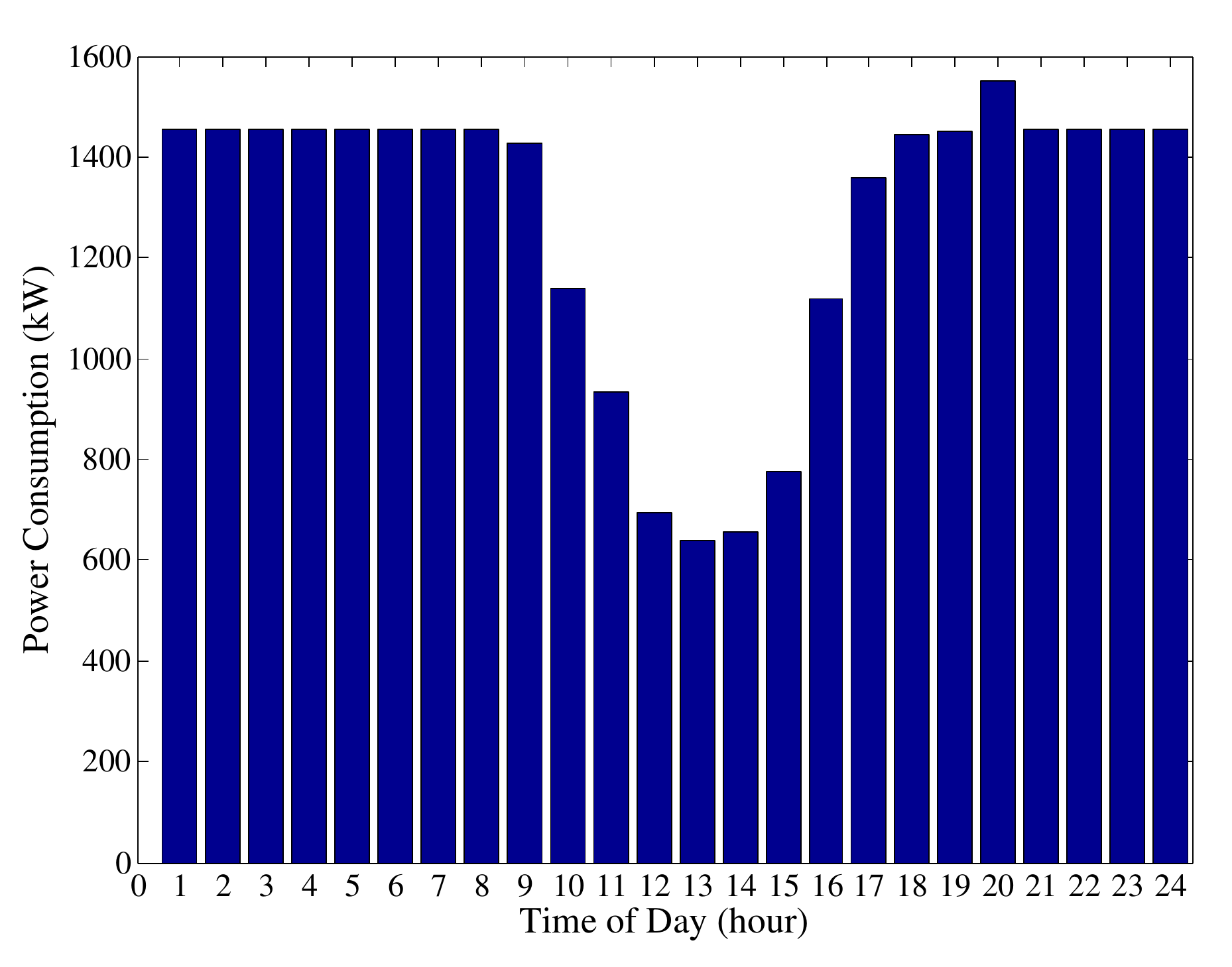}\vspace{-0.4 em} 
    \caption{Aggregated optimized daily power demand profile}
    \label{optvar}
    \end{subfigure}
         \begin{subfigure}[b]{\columnwidth}\centering
        \includegraphics[height=2.018in]{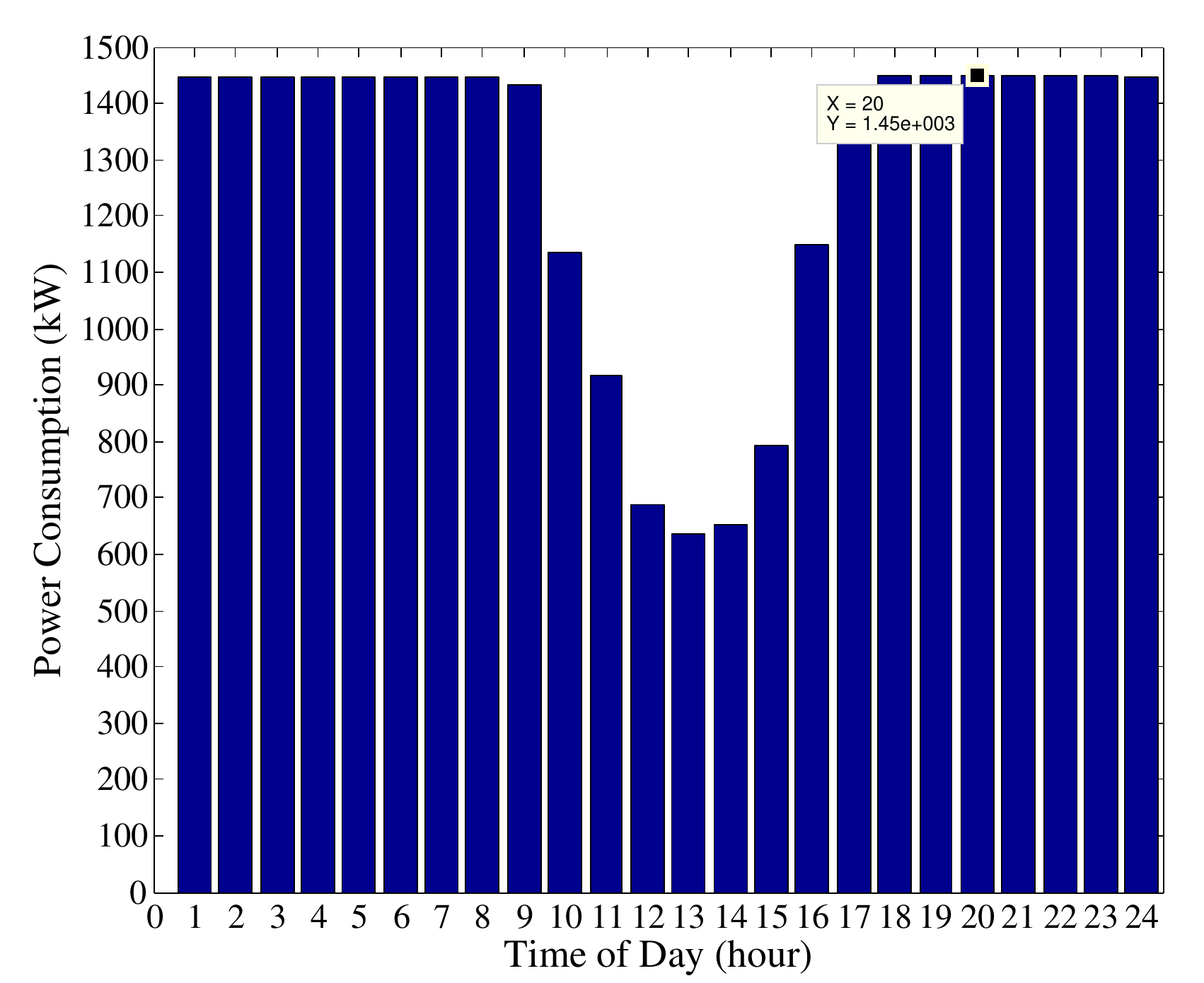}\vspace{-0.4 em} 
        \caption{Aggregated optimized daily power demand profile with V2G}
        \label{optvarV2G}
        \end{subfigure}
 \caption{Aggregated daily power demand profiles.} 
 \label{cpowprof}
\end{figure}

The results for the expected daily power demand of a typical EV under the aforementioned settings are illustrated in Fig. \ref{com}. As can be observed, the expected daily power demand resulting from the charging time distributions which possess the same mean and variance tends to the same power profile. However, for the exponential distribution, we see that its expected power demand differs significantly from that of the others. This gives us useful insights about charging power demand expectation of an EV throughout a day.

Next, we examine the DR scheme introduced in the previous section. Thus, we consider 1,000 users and the scheduling horizon is set as 24 hours. The users' EVs' arrival time and charging time are the same as the previous subsection. For the EV's departure time, a Normal distribution with mean at 7:30 am and one hour standard deviation is considered. Since the departure time must be greater than or equal to the arrival time plus charging time for each user, the \textit{accept-reject} method is employed here again. Fig. \ref{cpowprof} shows the aggregated demand profile of this system for four different cases: a) when there is no EV in the system, b) when EVs are in the system but their charging is uncoordinated, c) for coordinated charging of EVs according to (\ref{cor}) and d) when vehicle-to-grid (V2G) energy delivery is also enabled in the system. As it can be observed in Fig. \ref{optvar},  DR technique introduced in the previous section not only results in a more flat daily profile but also in a less peak demand which happens at 8:00 P.M. and is equal to the peak of the inflexible power demand profile shown in Fig. \ref{inflex}. This peak value can be decreased further by 100 kW when we have V2G capability in the system as shown in Fig. \ref{optvarV2G}.

\section{Conclusion and Future Work} \label{Con}

In this paper, we first presented a model for uncoordinated charging power demand of EVs based on a stochastic process and then we characterized an EV's expected daily power demand profile. Next, we illustrated an EV's expected daily power demand profile for different charging time distributions through simulation. We observed that large-scale accommodation of EVs with uncoordinated charging demand can significantly change the daily energy demand profile. We then introduced an autonomous demand response (DR) to manage the EVs' charging demand. Our results show that, it is possible to accommodate a large number of EVs and yet achieve the same peak-to-average ratio (PAR)  as when there is no EV in the system. We also showed that this peak value can be decreased further significantly when the system has V2G capability. In our future work, we aim to investigate appropriate DR techniques to manage EVs random charging demands when multiple retailers compete in an energy market, where the objective is to minimize the cost of the electricity provided by retailers to their corresponding.

\appendix

\begin{proof}[Proof of proposition \ref{pro}]
Since $x(t)=0$ for $t_0\leq t-T$ and $t\leq t_0$. Then, $\mathbb{E}[x(t)]$ becomes:  
\begin{gather}
\nonumber \mathbb{E}[x(t)]=a\times P(t_0 \leq t \leq t_0+T)\\ 
=a\times P(t-T \leq t_0 \leq t).  
\end{gather}
Further, we can use the \textit{total probability theorem} to get
\begin{gather}
\nonumber \mathbb{E}[x(t)]= a\times\int\limits_0^\infty P(t-T \leq t_0 \leq t | T=T') f_T(T')dT'\\
 = a\times \int\limits_0^\infty (F_{t_0}(t) - F_{t_0}(t-T')) f_T(T')dT' \\ 
 =a\times \bigg[F_{t_0}(t)-\int\limits_0^\infty F_{t_0}(t-T')f_T(T')dT'\bigg]
\label{int}
\end{gather} 
for which we have taken into account the facts that $\int\limits_0^\infty f_T(T')dT'=1$, and $t_0$ and $T$ are independent. Furthermore, we can express (\ref{int}) in a more concise form by using the definition of the convolution integral and the identity $f(t) \ast \delta(t)=f(t)$ as follows:   
\begin{align}
 \mathbb{E}[x(t)]=a\times \left( F_{t_0}(t) \ast [\delta(t)-f_T(t)]  \right).
\end{align}
\end{proof}

\bibliographystyle{IEEEtran} 
\bibliography{IEEEabrv,myBIB}

\end{document}